\documentclass[prl, amsfonts, twocolumn, nofootinbib, showpacs]{revtex4}
\usepackage{graphicx, epsfig}

\newcommand{\be}{\begin{equation}}
\newcommand{\ee}{\end{equation}}
\newcommand{\bea}{\begin{eqnarray}}
\newcommand{\eea}{\end{eqnarray}}

\newcommand{\gapp}{\mathrel{\raise.3ex\hbox{$>$}\mkern-14mu
              \lower0.6ex\hbox{$\sim$}}}
\newcommand{\lapp}{\mathrel{\raise.3ex\hbox{$<$}\mkern-14mu
              \lower0.6ex\hbox{$\sim$}}}
\def\bbox{{\,\lower0.9pt\vbox{\hrule \hbox{\vrule height 0.2 cm
\hskip 0.2 cm
\vrule  height 0.2 cm}\hrule}\,}}
\begin{document}
\title{Why black hole production in scattering of cosmic ray neutrinos
is generically suppressed}

\author{Dejan Stojkovic$^1$}
\author{Glenn D. Starkman$^{1,2}$}
\author{De-Chang Dai$^1$}
 \affiliation{$^1$ Department
of Physics, Case Western Reserve University,
             Cleveland, OH~~44106-7079}
\affiliation{$^2$ Astrophysics Department, University of Oxford,
Oxford, OX1 3RH, UK}

\begin{abstract}
 \widetext
It has been argued that
neutrinos originating from  ultra-high energy cosmic rays  produce
black holes deep in the  atmosphere in models with TeV-scale quantum gravity.
Such black holes  would initiate quasi-horizontal showers
far above the  standard model rate, so that the Auger
Observatory would observe  hundreds of black hole events.
This would  provide the first opportunity for  experimental study of
microscopic black holes.
However, any  phenomenologically viable model with a low scale of quantum gravity
must explain how to preserve protons from rapid decay mediated by virtual black holes.
We argue that unless this is accomplished by the gauging of baryon or lepton number,
the suppression of proton decay will also suppress quantum gravity
mediated lepton-nucleon scattering, and hence black hole production by
scattering of ultra-high energy cosmic ray neutrinos in the atmosphere.
 We discuss explicitly the split fermion solution to the
problem of fast proton decay.
\end{abstract}
\pacs{???}
 \maketitle

It is widely accepted that black holes can be produced
in collisions of particles with center-of-mass (COM) energy $E_{CM}$
larger than the scale $M_*$ at which
the quantum mechanical nature of gravity is manifest.
If two incoming particles collide with $E_{CM}>M_*$,
and an impact parameter smaller than the gravitational radius corresponding to $E_{CM}$,
then a black holes with mass $E_{CM}$ forms.
This is of little practical interest in theories where
quantum gravity is manifest only at the Planck scale,
$M_{Pl}\simeq 10^{19}$GeV.
However, the quantum gravity scale could be as low as $1$TeV \cite{bwm}.
The exciting  possibility is that black holes could then
be produced and studied
in near-future accelerator experiments \cite{acc}.
For  example, the Large Hadron Collider (LHC),
due to start operating in  $2007$,
will have $E_{CM}\simeq14$TeV.
Numerical estimates show that it
should be able to produce ${\cal O}(10^7)$ black holes per year
if $M_{\star}\lapp1$TeV.

In \cite{FengShapere}, it was pointed out
that black holes could also be produced by ultra-high energy cosmic rays
interacting with nucleons in the atmosphere,
with $E_{CM}>100$TeV.
 These microscopic black holes would
 decay nearly instantaneously through Hawking radiation,
producing extremely energetic cosmic ray showers.

The authors of \cite{FengShapere} argued that cosmic neutrinos with energies
above $10^6$GeV are effective sources of black holes,
with production cross-sections large enough to be relevant
for near-future cosmic ray observatories and well understood fluxes.
Neither strong nor electromagnetic interactions degrade the neutrino energy
before it interacts quantum-gravitationally, and,
since the neutrino interaction length is far longer than the thickness of the Earth's atmosphere,
neutrinos produce black holes uniformly at all atmospheric depths.
Therefore, the most promising signal for black hole creation by cosmic
rays should be quasi-horizontal showers initiated by neutrinos deep in the atmosphere.

It is straightforward to estimate the cross-section
for production of a $(4+n)$-dimensional black hole \cite{acc}.
Consider two particles moving in opposite direction with $E_{CM}=\sqrt{\hat{s}}$.
If their impact parameter is smaller than
the Schwarzschild radius of a $(4+n)$-dimensional black hole of mass $E_{CM}$,
\be
R_S \approx {1 \over  M_* } \left( { M_{{\rm BH}} \over M_*} \right)^{\frac{1}{1+n}}  \  ,
\ee
then a black hole with mass $M_{{\rm BH}} =\sqrt{\hat{s}}$ forms.
Thus, the black hole production cross-section is estimated to be
the geometric cross-section of the resulting black hole,
\be \label{gcs}
\hat{\sigma} (ij \rightarrow {\rm BH}) \approx \pi R_S^2  \, ,
\ee
where $i$ and $j$ denote the two colliding particles.
The total black hole production cross-section in neutrino-nucleon scattering is thus
\be \label{tcs}
 \sigma (\nu N \rightarrow {\rm BH} ) \simeq
\sum_i \int_{M_{min}^2/s}^1 dx \, \hat{\sigma}_i(xs)f_i(x,q) \, ,
\ee
where $s=2m_N E_\nu$.
($m_N$ is the nucleon mass and $E_\nu$ is the neutrino energy.)
The sum runs over all partons in the nucleon,
with $f_i$ the parton distribution functions.
$M_{min}$ is a minimal mass for which this formula is applicable ($M_{min} \sim M_*$),
while $q$ is momentum transfer.
The cross section for black hole production is found
to be several orders-of-magnitude higher than the standard model cross-section
for $\nu N \rightarrow L X$, if $M_{min} \approx 1-10$TeV.

Quasi-horizontal showers in the atmosphere can be observed by air
shower ground arrays or air fluorescence detectors. The largest current
and near-future cosmic ray experiment is the Auger Observatory.
Numerical
estimates using (\ref{tcs}) indicate that hundreds of such black holes
events may be imminently observed at the Auger Observatory in its first
five-year run ($2003-2008$) \cite{FengShapere}, before LHC data becomes
available.

While the production of quantum black holes seems inevitable in
particle collisions at center-of-mass energies above the quantum gravity scale,
there are important caveats arising from the stability of the proton.
The proton lifetime is greater than $2.1\times 10^{29}$ years \cite{pdb}.
More particularly it is $>10^{33}$ years in the channel $p \rightarrow e^+\pi^0$,
and $>10^{32}$ years in the channel $p \rightarrow \nu\pi^+$.
(The latter limit is technically from neutron stability inside nuclei;
the direct proton limit is an order of magnitude weaker.)
In principle, baryon and lepton number
violating operators that connect quarks and leptons,  of the form QQQL
(where Q denotes a quark, while L denotes either a charged lepton or
neutrino), can mediate proton decay.  Such operators are expected in
both grand unified theories and quantum gravity. Since protons are the lightest
baryons, and since leptons are the only free fermions lighter than
protons, baryon and lepton number  conservation would be violated
(assuming electric charge and fermion number are conserved) if a proton
could decay.

Baryon and lepton number violating operators do not occur in the
perturbative standard model.  However, these conservation laws follows
 from
an accidental low energy effective global symmetry, in the sense that
all the renormalizable operators that are consistent with the gauge
symmetry of the electroweak standard model turn out to conserve baryon
and
lepton number as well.  B and L (thought not B-L) are violated
non-perturbatively in the standard model.
B and L can also be violated at high energy. For example, quantum
gravity is expected to violate all  conservation laws that follow from
 such
effective global symmetries \cite{Hawking,AKP,SSA}, nor are they protected by
claims of unitarity in black hole evolution \cite{SSA}.  In particular,
quantum gravity mediated processes  like
\be p  \rightarrow \pi^0 + e^+ ,\ \  p \rightarrow \pi^+ + \nu \ \ {\rm
and} \ \
p  \rightarrow \pi^+ + {\bar \nu} \
,
\ee
would be expected to give the proton a life-time \cite{Hawking,AKP}
\be \label{plt}
\tau_p \sim m_p^{-1} \left({M_{*}}/{m_p}\right)^4  \, .
\ee
($m_p$ is the proton mass.)
 If $M_*=M_{Pl}=10^{19}$GeV we have $\tau_p
\sim
 10^{45}{\rm yr}$, much longer than the current limits. In
models with TeV-scale
 quantum gravity, i.e. $M_*=10^{3}$GeV, one loses $64$ orders of
magnitude in (\ref{plt}).  This results in a disastrously short proton
life-time.

Consider the process in which a baryon, $B$, such as a proton,
decays to a anti-lepton, ${\bar L}$,  such as a positron,  or
anti-neutrino,
plus other particles (which we denote as X) that carry no net B or L
quantum numbers,
via some intermediate quantum gravity state BH
(which we shall call a black hole despite any subtle technical issues),
\be\label{BtoLdecay}
B \rightarrow {\rm BH} \rightarrow  {\bar L} +  X  \  .
\ee
The possibility of (\ref{BtoLdecay}) implies the possibility of
\be\label{BLviolating-scattering}
B +  L \rightarrow {\rm BH} \rightarrow  X.
\ee
and {\it vice versa}.
In (\ref{BLviolating-scattering}),
the lepton scatters off a baryon and produces a menagerie of particles, X,
mediated by the same intermediate quantum gravity state.
These same arguments apply to ($(B-L)$-violating) proton decays to leptons,
implying the possibility of $B{\bar L}$ scattering.

To stabilize the proton,  one must somehow forbid (\ref{BtoLdecay}).
There are two approaches. One can prevent the formation of the
intermediate black hole, or one can ensure that the black hole
decays back to a state with the same baryon or lepton number as the
initial state.  In order for the black hole decay to depend on the
properties of the pre-existing initial state, either the black hole
or its environment would itself need to carry the information about
the quantum numbers of the initial state.

For the black hole itself to remember the initial pre-black-hole
initial state, it must carry B or L. This requires that we promote
the U(1) invariance associated with B or L  from a global to a local
symmetry --
{\it i.e.} 
that we gauge $U(1)_B$ or $U(1)_L$ invariance.   If B  is
gauged then, although the the B-violating processes (\ref{BtoLdecay})
and
(\ref{BLviolating-scattering}) are prohibited, the B  conserving
processes
\be\label{Lviolating-scattering}
B  + L \rightarrow {\rm BH} \rightarrow  B  + X,  \quad {\rm and}
\ee
\be\label{BLconserving-scattering}
B  + L \rightarrow {\rm BH} \rightarrow  B  + L  + X
\ee
are permitted.
Similarly, if L  is gauged then, although 
(\ref{BtoLdecay}) and (\ref{BLviolating-scattering}) are prohibited,
the L conserving processes
\be\label{Bviolating-scattering}
B  + L \rightarrow {\rm BH} \rightarrow  L  + X
\ee
and (\ref{BLconserving-scattering}) are permitted.  Finally, even if
both B and L are gauged,
the B and L conserving process (\ref{BLconserving-scattering}) is
allowed.

Gauging B or L permits cosmic ray neutrinos scattering off
atmospheric nucleons to produce black holes and yet protects the proton
 from decay. However, gauging B or L has proven to be problematic.
If $U(1)_B$ were an unbroken gauge symmetry,
there would be a long range interaction not seen in experiments.
Therefore,  $U(1)_B$ needs to be broken down to a discrete gauge symmetry.
The leftover discrete symmetry can preserve baryon number modulo
some integer \cite{KraussWilczek}. To suppress dangerous $ n
\rightarrow \bar{n}$ oscillations \cite{nnbar} one must forbid both
$\Delta B = 1$ and $2$ operators. The lowest allowed operators would
then be $\Delta B = 3$ (which are of dimension $12$ and higher). The
most common problem is arranging for cancellation of gauge anomalies
\cite{IbanezRoss}. This requires either an unusual charge assignment
to existing particles or the existence of new exotic particles.
There are other problems related to the idea of gauge couplings
unification. Although one cannot exclude gauging $U(1)_B$ or
$U(1)_L$, one must allow for the possibility that they are not, and
that the stability of the proton in low-scale quantum gravity models
is due to other causes.

The second possibility is for the black hole's environment
to preserve the information about the pre black hole initial state.
The only thing that distinguishes the place where the black hole was
created from any other is that it is on the brane.
Thus for the environment to inhibit B or L violating black hole decay,
there must be different locations in the bulk for baryons or leptons.
This is precisely the split fermion model \cite{ArkaniHamedetal}.
We shall find that in this model, black hole production is heavily suppressed.

As an alternative to gauged baryon number,
\cite{ArkaniHamedetal} proposed
that standard model fields are confined to a ``thick'' brane -- much
thicker than $M_*^{-1}$. Quarks and leptons are stuck on different
three-dimensional slices within the thick brane (or on different
branes), separated by much more than $M_*^{-1}$. This separation
causes an exponential suppression of all direct quantum-gravity
couplings between quarks and leptons, due to exponentially small
wave functions overlaps. The virtual black holes associated with
protons therefore have an exponentially small probability of
decaying with emission of leptons \cite{HKM}, and thus of mediating
lepton number violation. Assuming that black holes do not mediate
electric charge non-conservation or fermion-number violation, this
stabilizes the proton. For appropriately chosen spatial separation
$d$ between the quarks and leptons, the proton decay rate can, as we
shall see,  be safely suppressed.

The propagator between fermions which are separated in the extra
dimension was derived in \cite{ArkaniHamedetal}:
\be \label{p}
P_{d} (t)  \approx  \left({\sqrt{2\pi} R}/({\sigma t })\right)
\exp\left[-d^2/(2\sigma^2)\right]  ,
\ee
$R$ is the characteristic size of the extra dimension
($d\sim R$)
and $\sigma$ is the inverse width of the fermion wave function in the extra dimension.
The momentum transfer of $t$-channel scattering is $\sqrt{-t}$.
The expression in (\ref{p}) is valid in the limit of large momentum transfer,
$\sqrt{-t} \gg {\rm max}\left(R^{-1}, d/\sigma^2\right)$.
The propagator has the usual four-dimensional momentum dependence save
that the coupling is suppressed by the exponentially small wave function overlap.
The proton decay width is proportional to the propagator squared.
To adequately suppress proton decay,
we need to increase the lifetime by ${\cal O}(10^{51})$.
This implies that $e^{-d^2/(2\sigma^2)} \sim 10^{-25.5}$,
or a rather modest hierarchy $d \sim 10 \sigma$. It is obvious that
black hole production via such direct tree level quark-lepton interactions 
will be heavily suppressed.

What about higher order effects? Gauge bosons mediate lepton-quark
interactions, so might enable proton decay and black hole
production. The inability of gauge bosons to carry fermion number
between branes prevents their contributing to proton decay, but not
to black hole production.

Gluons might be localized on the same brane as quarks in which case
there will be the same exponential suppression  of interactions with the
leptons, as in the tree level lepton-quark interactions. Alternatively,
the gluons may propagate freely between the quark and lepton branes.
Standard model interaction between leptons and gluons are of course suppressed
because leptons carry no color. However, leptons could, in principle,
interact directly with gluons via quantum gravitational interactions.
Similarly leptons could interact directly with the weak gauge boson ($W^\pm$ or $Z$)
or photon content of a nucleon to make a black hole.  (Or the gauge
boson content of the leptons
could interact with the quarks of the nucleon,  or finally the gauge boson
content of the leptons could interact with that of the proton.)
This too, however, is suppressed compared to expectations.

There are two separate limits to consider -- hard gauge bosons and
soft gauge bosons.  Although the hard gauge boson content of the
nucleon near the quark brane is  considerable, this falls rapidly as
one moves deep into the the thick brane. There is first of all a
volume suppression to the cross section, proportional to the ratio
of the lepton brane volume to the volume of the full bulk,
$(\sigma/d)^n$, where $n$ is the number of extra dimensions (for
realistic models one usually takes $n \geq 2$, however recent
experimental limits seem to force $n \geq 3$ ). There is a further
exponential suppression since the harder the virtual gauge bosons,
the closer to on-shell it must be. Only ones very nearly on-shell
can propogate the large distance between branes. However,
the gauge bosons are getting more massive  as they move into the
bulk (due to the effect of their Kaluza-Klein towers) and the
propagator from the plane where they live (with their bare masses)
into the bulk is highly suppressed \cite{ArkaniHamedetal}.

Soft gauge  bosons more readily negotiate the interbrane gulf, after
all they mediate standard model interactions, but  they are
insufficiently  energetic to contribute much to the center of mass
energy.  In order to make a black hole in a scattering they must
encounter a much higher energy  neutrino.  The neutrino flux is a
rapidly falling function of energy, so the rate of black hole
production by this mechanism is  further suppressed (in addition to
the above mentioned volume suppression factor).

The interaction between leptons and the Kaluza-Klein graviton
content of a proton faces the same issue as between leptons and the
protons's gauge boson content. And will be at least suppressed by
the same geometric factor that applies for delocalized gauge bosons,
i.e.  $(\sigma/d)^{n} \simeq 10^{-n}$.

So, the production of the small black holes whose radius is much
smaller than the quark-lepton separation will be suppressed. What
about larger black holes that can bridge the gap between the
fermions? Their production will be also suppressed. To see this,
note that the Schwarzschild radius of a higher dimensional black
hole grows very slowly with its mass, $R_s \propto M_{BH}^{1/1+n}$.
In order to overcome the large ($\sim 10 M_*^{-1}$) quark-lepton
separation, one requires COM energies larger by a factor of
$10^{n+1}$ than otherwise expected, and therefore incident neutrino
energies larger by $10^{2(n+1)}$. The neutrino spectrum at these high
energies is believed to be falling as $\sim 1/E^{3}$ which gives a
suppression of  $\sim 10^{18}$ for two extra dimensions, and much
more for more extra dimensions.

Thus, all the potential black-hole producing interactions are
suppressed by wave function overlap factors, whether exponential or
geometric. Black hole formation will thus be well suppressed in the
split-brane model.

\begin{figure}[h!]
    \centering{   \includegraphics[width=3in]{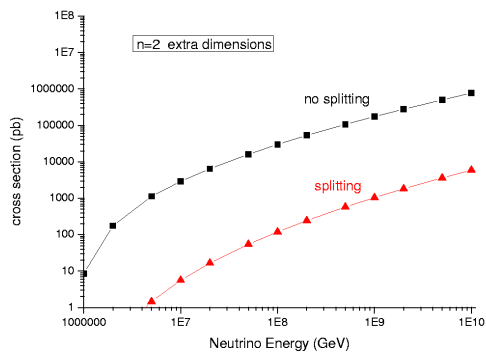} }
\caption{The black hole production cross section for the
conventional (non-split) and split fermion models. We take $n=2$
extra dimensions,
 the fermion wave function width $\sigma \sim M_*^{-1}$ and
 the separation between the quarks and leptons
 $d \sim 10 M_*^{-1}$, where $M_* = 1$TeV.}
    \label{ratio}
\end{figure}

As an illustration, in Fig. \ref{ratio}, we plot the black hole
production cross sections for the two cases (non-split and split
fermions). For the split fermion case, we use the equation
(\ref{tcs}), but instead of (\ref{gcs}) we use $\hat{\sigma} \approx
\pi (R_s^2 - d^2)$, where $d$ is the separation between the
fermions. Gauge bosons are allowed to propagate between the
fermions. This takes into account only a simple wave function
overlap suppression (no eventual loop suppression, energy lost  to
gravitational radiation etc.). Even in the most optimistic case of two extra
dimensions the suppression  factor is of the order of $10^3$,
in good agreement with the previous discussion. The suppression will
grow stronger with  more extra dimensions and with larger separation
between the fermions.

We note that the splitting of leptons from quarks would not suppress
$\Delta B=2$ processes, like $n-{\bar n}$ oscillations \cite{nnbar},
large left-handed Majorana masses for fermions, and
large mixing between the neutrino generations.
That would require, for example,
further splitting between up-type and down-type quarks,
or splitting between the different lepton generations.
However, the purpose of this paper is not to fix the split-fermion model,
but to use it as a simple explicit illustration of a model where
black hole production in lepton-nucleon scattering is suppressed.

We see that the correct black hole production cross section in
collisions of neutrinos and nucleons is not given by (\ref{tcs}).
Large suppression factors, ranging from wave function overlaps of
$10^{51}$ for direct tree-level processes, to more modest but still
substantial volume overlap factors  $10^{n}$ (with $ n\geq 2$, or
rather $ n\geq 3$ in the light of the latest experimental limits)
for gauge boson-lepton processes, divide the geometrical cross
section (\ref{gcs}) and thus enter the total production cross
section (\ref{tcs}) as well. This renders the corresponding
probability for black hole production by cosmic neutrinos
uninteresting for the Auger Observatory.

The final possibility is that production of the intermediate black hole
in (\ref{BtoLdecay}) is forbidden.
In this case, not only is (\ref{BtoLdecay}) suppressed, but so are all black-hole
mediated scattering processes between a proton and a lepton or anti-lepton.
We therefore learn that, in the context of low-scale quantum gravity,
unless the stability of the proton is due to the gauging of B or L,
neutrino-nucleon scattering (and antineutrino-nucleon scattering)
does not result in copious production of black holes.

We note that  black holes might still be produced
in atmospheric nucleon-nucleon or photon-nucleon scattering.
In the split fermion model, nucleon-nucleon scattering
arises because quark wave functions are not separated from themselves.
(Even if every flavor of quark was on a different brane,
the same quark flavor from both nucleons can scatter.)
Photon-nucleon scattering arises because the photons are not confined
to one or another brane. However, these process will lead to completely
different production cross sections and are much less certain.
(The flux of cosmic neutrinos used in
 \cite{FengShapere} comes from the decay of $\pi^\pm$ in collisions
 of ultra-high energy protons with the cosmic microwave background.)

Furthermore, the Earth's atmosphere is not  transparent to nucleons or
photons as it is to neutrinos.  The typical nucleon or photon  will
scatter by ordinary Standard Model processes many times before it
has a chance to scatter quantum gravitationally.
In so doing, it will typically lose too much energy to produce a black hole.
Only the rare photon or nucleon will scatter quantum gravitationally
early enough in the shower process to produce a black hole.
Moreover, even if the cosmic ray nucleon or photon flux,
and the scattering cross section are large enough to
produce detectable numbers of black holes,
the experimental signature would be much different than for those
created in neutrino-nucleon scattering.
In particular, one can not expect quasi-horizontal showers deep in the atmosphere.
Instead, these black holes would be produced high in the atmosphere.
The experimental signatures and expected rates merit further careful review,
but the prospects for detection are considerably less optimistic than
had been calculated for neutrinos.

The other possibility is to have black holes produced in
lepton-lepton (for example $\nu e^-$) scattering in the atmosphere.
From eq. (\ref{tcs}) we see that the total black hole production
cross section depends on the COM energy squared in the collision,
which in turn is proportional to the mass of the target. Since the
electron mass is $2000$ times smaller than the proton mass, the
threshold neutrino energy for leptonic black-hole production should
be $2000$ times higher.  This implies a significant reduction in
flux of neutrinos above threshold, since the flux goes down steeply
with energy ($\sim 1/E^{3}$). Since using the naive calculation of
neutrino-nucleon scattering one expected to detect a hundred black
hole events in a five-year run of the Auger Observatory, therefore
the expected suppression of $\approx 10^9$ in the neutrino-electron
case  renders the rate of black hole production completely
uninteresting in terrestrial detectors.

In conclusion, we reconsidered the question of black holes
production by cosmic neutrinos in models with TeV-scale quantum
gravity, originally considered in \cite{FengShapere}. We pointed out
that, unless baryon number or lepton number arise from gauge
symmetries, the stability of the proton suggests that lepton-nucleon
scattering does not lead to black hole production that might be
observable at the Auger. As an example, we showed that in the model
of split fermions, the rate of black hole production by cosmic
neutrinos is suppressed to an uninteresting level.  Although the
possibility of black hole production by high energy cosmic ray
nucleons or photons still remains, the fluxes are uncertain, and the
signatures and rates need to be properly reconsidered but are
pessimistic.

\vspace{12pt} {\bf Acknowledgements}:\ \
The authors are grateful to Tanmay Vachaspati, Francesc Ferrar, Jonathan
Feng, Alfred Shapere, Manuel Toharia and Luis Anchordoqui for very
useful conversations. The work was
supported by the DOE grants to the Michigan Center for Theoretical
Physics, University of Michigan, and to the particle astrophysics group at
Case Western Reserve University.

\end{document}